# Nikolaos Athanasios Anagnostopoulos

matr. nr.: 351698

# Exploring the complicated relationship between patents and standards, with a particular focus on the telecommunications sector

Strategic Standardisation

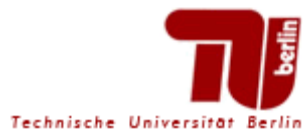

Technische Universität Berlin

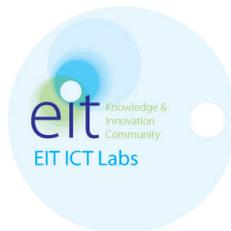

EIT ICT Labs Master School

*Abstract*

While patents and standards have been identified as essential driving components of innovation and market growth, the inclusion of a patent in a standard poses many difficulties. These difficulties arise from the contradicting natures of patents and standards, which makes their combination really challenging, but, also, from the opposing business and market strategies of different patent owners involved in the standardisation process.

However, a varying set of policies has been adopted to address the issues occurring from the unavoidable inclusion of patents in standards concerning certain industry sectors with a constant high degree of innovation, such as telecommunications. As these policies have not always proven adequate enough, constant efforts are being made to improve and expand them.

The intriguing and complicated relationship between patents and standards is finally examined through a review of the use cases of well-known standards of the telecommunications sector which include a growing set of essential patents.

# Introduction

A standard is a voluntarily agreed norm, a set of rules or definitions, usually describing methods and processes for producing or doing something. Standards are most often in the form of published documents setting out specifications and procedures designed to ensure products, services and systems are safe, reliable and consistently perform the way they were intended to.[1] Naturally, a standard aims to gain as much acceptance from the relevant industry as it is possible and to be implemented widely by it. Therefore, standards are usually formed using consensus decision-making, seeking the consent of the vast majority of all the participants of the standardisation process.

A patent is a form of protection of an invention, which could be a new process or device, or even a new material. A patent confers exclusive rights to inventors, or any people they may assign, on their invention, regarding its production, use or sale, in exchange for the public disclosure of this invention. A patent is awarded on a national level,[2] thus an inventor has to apply in each different state to have his rights recognised with the award of a patent.

Obviously, most parties are interested in creating a market for their inventions; this being a reason for them obtaining patents on their inventions. Slowly, these markets may gain some momentum and attract more people, or even huge companies, doing research and development on relevant fields. This often leads to the expansion of the market and, of course, to a growth of relevant patents being awarded to different vendors and manufacturers. This effect in turn leads to a need for standardisation so as to maximise efficiency, minimise cost and, naturally, also, maximise profit.

Various reasons may act as incentives for patent owners to take part in a standardisation process. One of the main reasons would, obviously, be to be able to influence the standardisation process, perhaps, while also promoting the inclusion of their patents instead of ones owned by their competitors. A patent owner needs to take part in the standardisation process even if it holds a major share of the market. Otherwise, its patents will probably be precluded in favour of others and, by the network effect and other economic effects which a standard creates, it is bound to lose a significant share of the market to competition.[3]

Another interesting incentive for patent owners to take part in the standardisation process is the fact that usually a major standard in a highly innovative sector relies on many different patents to be formed.[3][4] Thus, through the standardisation process, patent owners have the opportunity to normalise the licensing and cross-licensing processes and move towards a point that better serves the interests of all major parties. In essence, the standardisation process serves also as a negotiation process for patent owners to normalise the regime that governs the licensing and usage of their patents to their common benefit.[3]

Through the formation of a standard, patent infringement of its essential patents is bound to minimise, while the number of vendors and manufacturers which pay licensing fees for these patents will maximise.[3] Therefore, even though the inclusion of a patent in a standard may lead to lower licensing fees applying for it, it will also lead to more profit being made out of this patent, as the number of its licensees will maximise.

As it has been analysed by Swann's studies on the economics of standardisation,[5][6] standardisation can have a positive effect on a market by raising network externalities through compatibility standards, increasing the variety of products and enhancing the efficiency of supply chains through interoperability and interface standards, while also reducing transaction costs and employing economies of scale by variety reduction and, finally, facilitating trade by the commonly agreed rules and procedures it establishes.

In a market driven by different inventions and patents, unavoidably, almost any relevant standard will include some patent or other.[3] This is extremely prevalent in such sectors as telecommunications, which are driven by a high rate of innovation and thus any attempt for standardisation in them involves a great number of patents, usually owned by different participants.[3][7][8] The process of standardisation may therefore lead into conflicts or even standards wars.[9][10] However, as shown by the tree metaphor used in Swann's study,[5] standardisation is both beneficial and inevitable for such a market.

The mere presence of patents inside standards should cause no conflicts on its own, provided that these patents are indeed essential for the standard and the patent holder waives its rights, or makes licences available at a reasonable fee to all interested parties.[3][4] Problems arise if a patent included in a standard may not be deemed as truly essential for it, if two parties hold patents for different implementations of the same product, or if a party avoids licensing its included patents at a reasonable fee to all interested parties under the same criteria.[9][11]

Most standard setting organisations have adopted sets of rules which aim to prevent such conflicts. Such rules prevent the inclusion of any patents in standards other than those truly essential for them, those without which the implementation of a product, a function or a process is impossible and also regulate how those patents have to be licensed. Other rules, concerning the case where multiple patents exist, each one regarding a different implementation of the same general product, function or process, request the inclusion of only the patent which concerns the implementation which is considered optimal for the specific product.[12]

However, parties involved in standardisation may try to abuse this process for their own benefit by either hiding their ownership of relevant patents, manipulating the licensing process or even refusing to give a license.[3][11] Moreover, they try to form alliances and strategies which will lead into the inclusion of their own patents in a given standard and in the exclusion of the ones owned by a competitor, regardless of which patents are truly essential for the standard or provide a better implementation.[3][8]

This may give rise to the existence of competing, usually informal or regional, standards for the same general product, which lead to standards wars,[9] or may even undermine the whole standardisation process, preventing the adoption of a specific, usually formal, standard.[10]

# The relationship between patents and standards

As it has already been noted, patents and standards are both essential for promoting innovation and market growth.[7] However, it has also been argued that their goals may be seen as contradictory.[3][11][13] A standard, as defined by standardisation organisations,[14] is a document intended for public use and is obviously basing its success or failure on whether the rules, guidelines or techniques it provides are adopted by the majority of manufacturers and suppliers of the market or industry field it tries to address. A patent, on the other hand, is a document primarily asserting private rights on a new invention, whether that is a new process, technique or product. It usually does not primarily target so much on the right it offers to its owner to provide a licence to a third party, as much as on the right its owner has to exclusively produce, sell or even use this product or process. It aims to preclude others from the market, using the right its owner has to *not* provide a licence to any other third party, apart usually for the state's government, where the patent has been granted.

Therefore, the basic conflict between a patent and a standard is that a standard bases its success on the very thing a patent tries to avoid, widespread use of it by the public.[3][4] This difference, while true in its general case, may not always be as evident. A private (or company) standard holds many similarities to a patent as its, at least initial, intended use is very limited. In the same sense, a patent which its owners have declared they will licence to everyone for no fee, or for a negligible fee, shares a lot with the basic idea behind a standard. These extreme cases demonstrate that, in their very essence, a standard is a set of approved rules, guidelines and techniques, while, a patent is a description of a product or process. In really extreme cases, a patent and a standard, though produced by very different processes, may essentially account for the same thing. Generally, however, their intended uses are in conflict with each other.

For this reason and also because different, or even similar, techniques, methods and processes exist to perform similar, or even same, functions or produce similar products, including a patent in a standard can lead into major conflicts during the standardisation process or even to its complete collapse.[11]

An easily explained example of this would be a standard about a product made from four separate parts. There are a few different ways to assemble such a product, if the order in which parts are assembled does not affect the final product. But, assuming one company holds a patent (Patent A) on the process of assembling first two different sets of two parts with each other and then combining them into the final product, and another company holds a patent on assembling the parts one by one in a row (Patent B), the criteria with which to choose which assembling process is optimal and which patent, if any, should be included in the standard may not always be as easy to define as they seem.

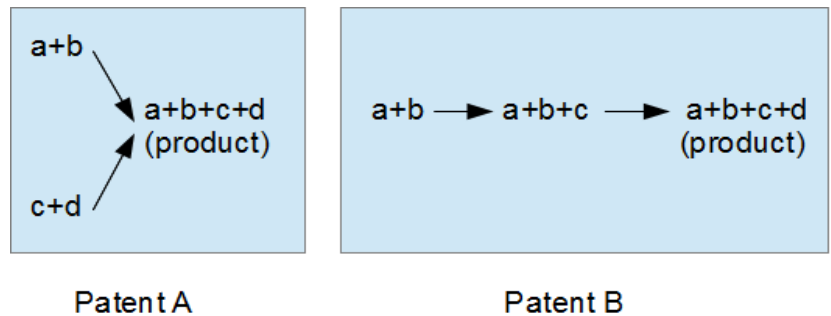

*Fig. 1: Patented processes A and B*

## Problems arising from the inclusion of patents in standards

It may appear easy to choose Patent A over Patent B, but the first thing that has to be determined is whether including a patent for such a process is essential for the standard. This is not always as evident as it seems. A company that holds a patent on such a process may try to have it included in the standard even if it is not really essential using common sense. It relies, however, on a technical committee, or, in general, the parties participating in standardisation to determine what is essential for the standard and what is not. Therefore, a company which has a major share of the relevant market, or which has significant influence on the other parties of the standardisation process may have its patents included in a given standard, even if those are not truly or uniformly recognised as essential for it.[4]

Even worse, a patent owner may push for the adoption of a certain process for which it holds a patent, without first informing the other parties for the existence of this patent, which would constitute a "non-disclosure strategy" for its standard-essential patents.[4][7][11][15] In such an event, and after the standard has been adopted, it's really difficult to address the resulting problems, as this may mean starting the standardisation process again from scratch. This could also occur with the patent of a party which was not taking part in the standardisation process, making the resolution of such an issue even more difficult.[7]

However, assuming a patented process has been legitimately and correctly identified as essential for a standard, it is still quite difficult to determine the right approach to this issue. The parties of the standardisation process may decide to try to circumvent the relevant technology in some way.[16] This again may prove difficult or lead to litigation over alleged patent infringement, which again may either hinder the standardisation process or result to its work having to be reconsidered from scratch.

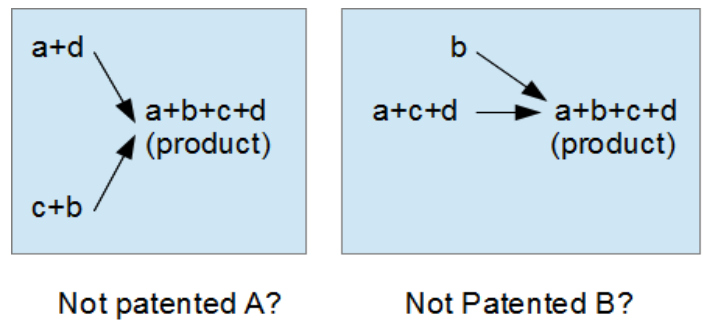

*Fig. 2: Trying to circumvent patents A and B*

If it is decided that a patented procedure or technique is essential for the standard and should be included, it again has to be determined which of the available patents over this or similar procedures and techniques is the optimal one to include. This can prove really hard to attain. Using again the example of patents A and B, patent A may seem to be offering a better solution than patent B. However, the actual time required by each patented process, their availability, their current adoption in the market, transition costs that may be required to move from the implementation of one process to the other, and, of course, the licensing fees that each one requires have to be taken into account. All these factors, combined with the position that each patent owner holds in the relevant market and the strategic alliances and partnerships it may share with other parties of the standardisation process may lead into the inclusion of a patent that may indeed be essential, but, doesn't seem optimal using common sense.[3]

This can become evident if we consider the fictional case of a standard about common everyday cars where one company holds a patent for regular car tyres and another for a car on rails. If the patented rail technology is widespread and holds an overwhelming percentage of the (regional) market, perhaps because of the flat terrain of the region and the economies of scale that have prevailed, whereas the car tyre technology is patented by a small company with limited production, it is quite likely that the (regional) standard will be one which includes the rail technology patent.

Although this scenario is definitely unimaginable about cars as we know them today, it clearly demonstrates some of the actual dynamics of the standardisation process. Obviously, a patented technology is essential in this case, as a car has to have wheels, and obviously the patented technologies are far better than plain wooden wheels. However, even if the small company that owns the car tyre patent is somehow notified and takes part in the standardisation process, it will be really difficult to have its own patent included in the standard because of the prevalence of rails in the market. Only if the market shifts for car tyres, will this particular standard start to include the relevant patent.

Regional, informal, or even *de facto* standards have been formed with the same rationale as the previous example. A standard, unlike a patent, aims for the highest possible consensus, rather than the best or most optimal innovation. Thus, in sectors like telecommunications which are driven by a high rate of competitive innovation and conflicting patents, it is quite usual to also have incompatible regional or informal standards, where each one includes one or more patents that perform similar, or even the same function, but are owned by different parties and implement similar, but yet different, processes and techniques to do so.[3][4]

Moreover, as standardisation can be a long and difficult process, entities taking part in it may be granted patents during or even after the standardisation process which may include patents for parts which have already been assessed as essential for the standard under consideration. This may lead to patent ambush, where the patent holder suddenly asserts that one of its patents is, or will be, infringed by use of the standard, while not having done so before or during the appropriate phase of the standardisation process.[7]

Finally, a party may move to acquire more relevant patents during the standardisation process to have a better negotiating position during it or push for inclusion of more of them in the standard as essential.[3] It may also do so to attain rights on technology that it has identified as more innovative or having more potential than the one it already holds rights upon.

For example, the patent owner of the 'car on rails' technology may attempt to buy the 'car tyres' technology during or even after the standardisation process, as it has identified during standardisation that doing so will require a quite low cost in return for a greater future profit, especially if the 'car tyres' technology has not been selected for inclusion in the standard.

# Proposed and adopted solutions concerning the problems arising from the inclusion of patents in standards

Most major standardisation organisations seem to have adopted rules regulating both the terms under which a patent may be included in a standard, but also highlighting an obligation of all parties of it to declare their possession of any relevant patents, even if those have not yet been explicitly identified as essential for the standard.[7][17][18][19][20]

Rules have also been set concerning the terms under which essential patents have to be licensed. More and more standardisation setting organisations adopt what has come to be called the "F/RAND obligation" for their members, an acronym which stands for the obligation of the parties participating in their standardisation processes to license their essential patents under "fair, reasonable and non-discriminatory" terms. However, such obligations have not been uniformly or officially adopted by all standardisation bodies, usually to allow for more flexibility in both standards and patent licensing negotiations. In spite of the exact terms they use, all major standardisation organisations have adopted some policies identifying and addressing a need for standards, and therefore also essential patents, to be accessible to everyone without undue constraints.[7][17][18][19][20] Moreover, the "F/RAND obligation" have been constantly gaining ground not only as adopted rules for the standardisation process, but also as preferred rules governing litigation that concerns issues with essential patents.[21]

It can be argued that in a standardisation process which involves patents which have been claimed, identified or accepted as essential, a form of negotiation between all parties concerned takes place. While this discussion may initially concern the very inclusion of a patent in the standard, when or even before consensus has been achieved on this issue, a more interesting negotiation commonly takes place over the licensing terms and fees of such patents. These negotiations are usually governed by the "F/RAND obligation" or similar rules imposed on these patent holders who take part in the standardisation process, and could thus lead into licensing fees being lowered. However, as it has already been explained, the inclusion of a patent in a standard, will rather increase the patent owner's total profit, as a licence for using an essential patent will have to be acquired by every user of the standard.[3]

An additional interesting side effect of the inclusion of a certain patent in a standard as essential for it is that it renders all other competing patents rather valueless, except if those are included in other competing standards. Even then, at the time when a particular standard prevails over its competing ones, usually, all its essential patents also prevail over any competing ones, rendering those essentially worthless.[22] That is another important issue explaining why patent owners have to participate in standardisation.[16]

Another way to prevent abuse of essential patents by patent owners is to have them disclose the existence and ownership of their relevant patents and their suggested licensing terms and fees before or during the standardisation process and before those are selected to be included in the final standard. Ex ante disclosure of licensing terms for patents is not, however, currently mandatory or widespread, but, it can particularly diminish the risk of patent hold-up, where a patent owner will demand a higher royalty rate than he could have negotiated when the patent was not essential for the standard.[7][15]

The most effective and widespread solutions to problems concerning the inclusion of patents in standards are cross-licensing mechanisms and patent pools.[7] Cross-licensing refers to the practice where two or more parties agree to license some of their patents to each other, while a patent pool is a group of patents regarding a particular technology that can be licensed together under common shared fees and terms. Patent pools of essential patents help avoid the risk of royalty stacking in a standard where multiple royalties for different essential patents have to be paid to different patent owners, while also avoiding the risk of a particular patent owner refusing to license his patents to certain parties or licensing them under discriminatory or unreasonable terms.[15][22]

Licensing under F/RAND terms is essential for the quick build-up of consensus and the adoption of a standard, as in the opposite case, the inclusion of certain patents may be blocked by those parties believing they are not being offered equally privileged terms as the others. In such a case, the standardisation process may even collapse or be radically delayed, until either the offending patents are somehow circumvented or an agreement is reached which will be considered satisfactory by all parties.[11] Cross-licensing is a common way for patent owners to resolve patent disputes.[22]

Additionally, licence contracts over essential patents may contain clauses aiming to secure the specified licensee will, even retroactively, get as favourable terms as the licensor may offer to any other licensee.[11]

Another, quite uncommon, possibility is that a patent owner of some essential patents may decide to license them for free or even place them on the public domain. This usually occurs when a company has decided to withdraw from a particular market or sector,[4] and can be explained as a move of good will, which may gain precious support, partnerships or alliances in other markets or sectors.

Yet another interesting detail is that not only standardisation organisations are concerned about the role of patents in standards, but, also intellectual property organisations and groups take an interest on how essential patents interact with standards,[13][23] while some of them have gone so far as to set their own rules for the inclusion of patents in standards.[13]

# Further complications concerning the inclusion of patents in standards and possible countermeasures

A set of different interpretations exists for the F/RAND terms: *fair*, *reasonable* and *non-discriminatory*. One suggested interpretation is that the patent owner of an essential patent for a given standard must license it to all other companies in the industry. Another interpretation is that the patent holder must license the patent to every other company in the industry under the same terms, whereas yet another different interpretation is that the patent must be licensed under similar terms to companies with similar circumstances.[11]

Other conflicts may arise from the interpretation of each individual word of the term "F/RAND obligation".[11][24][25][26][27] These differences in interpretation and the general vagueness of terms used by standardisation organisations in their rules may lead to litigation between parties of the standardisation process or sanctions by the national authorities responsible for the protection of consumers and the regulation of industry and market sectors.[8][11][15][25][26][28]

In some cases, technology that has to be included in a standard may be part of a larger bundle of patents containing also other technologies, which are irrelevant to the standard, or the patent owner may claim that he was not aware of owning an essential patent until after the standard has been adopted due to the size of his patent portfolio. Both of these cases usually lead to court litigation and sanctions.[11]

Furthermore, patent holders may refuse to provide a licence for some specific patents to certain vendors to prevent the customers of these vendors acquiring rights on those patents through the licence that may have been provided to these vendors. However, it has been argued additionally that the amount of licensing fees required for such patents may not suit the business strategies of these vendors either.[11]

Another strategy some patent owners may employ is a case of patent hold-up by attempting to raise the licensing fees they demand over patents generally relevant to a standard, some time before, in the beginning or during the standardisation process, as this would give them an advantage in negotiations over the licensing fees if any of these patents are selected to be included in the standard as essential for it.

Other issues include essential patents which are owned by third parties not participating in the standardisation process and thus not having an "F/RAND obligation" over the licensing fees and terms of these patents, and the difference between policies and rules of different standardisation organisations which may use different terms or even have completely different sets of rules concerning the inclusion of patents in standards and the relevant licensing negotiations. The latter situation may lead patent holders into forum shopping for the purposes of standardisation.

In general, it can be concluded that non-compliance with any disclosure or licensing rules, obligations and guidelines of a particular standardisation organisation over essential patents, or in general, will inevitably lead to court litigation between different parties resulting into possible sanctions and/or actions taken by national regulatory authorities. However, the ambiguity of terms and rules used by standardisation bodies is so extensive that neither courts nor regulation authorities have so far been able to completely normalise the rules and laws that concern standard-essential patents.

It has also been suggested that the current state of intellectual property rights databases, especially the one maintained by ETSI, the European Telecommunications Standardisation Institute, is not very transparent or complete,[7][15] and thus their figures are anything but reliable, leading into confusion over whether the patents included there are all truly essential for their related standards and, also, over their ownership status and their licensing terms and fees.[29][30]

Some of the suggested solutions to these complications faced by the inclusion of patents in standards may amount to compulsory ex ante disclosure of all patents considered relevant to a standard and potential compulsory licensing of those of them that are deemed essential for the standard.[11][15][24] Other countermeasures that have been proposed include assuring a high level of quality of essential patents, which would reduce the risk of conflicts arising from low quality patents, and also promoting stronger collaboration between intellectual property organisations and standardisation bodies.[7] Another suggestion concerns the conciliation of competing standards with the participation of all parties concerned, when that is possible, resulting into a single standard being adopted and supported by all the parties.

It has also been suggested that worldwide harmonisation of national rules and laws concerning patents and standards should be achieved in order to decrease the likelihood of conflicts caused by cross-border application of technical standards. Further countermeasures that have been suggested include improving the transparency and accessibility of material relating to patents, such as European and foreign case law on intellectual property and competition policy rulings, and decreasing the cost of monitoring patents in standards, by improving and assuring the transparency, completeness and actuality of the intellectual property rights databases.[7]

Moreover, there has been a call for rules that would guarantee the timely and precise disclosure of essential patents, would address the transfer of essential patents to third parties, would demand the irrevocable and worldwide applicable licensing commitment of patent owners for their essential patents and would ensure that already given licensing commitments are enforced.[7][24]

Finally, such rules as described above must be transparent and clear to all parties and their application should be uniformly coordinated, if not harmonised, between different standardisation organisations.[7][24] In conclusion, there has been a call for common clear rules adopted by all standardisation bodies which would address the identified shortcomings of the present situation. There have also been suggestions for stronger collaboration between intellectual property rights groups and standardisation organisations and for easier access to information concerning the existence and ownership of intellectual property rights and, also, to rules, court rulings and laws on intellectual property and competition policy.

# Complications encountered in the telecommunications sector

It has to be recognised that one of the most affected industry and market sectors by the inclusion of patents in standards, if not the most affected one, is the telecommunications sector. As telecommunications are characterised by a high degree of recent innovation and continuous market growth, a growing number of essential patents exist in this sector. It can be argued that standardisation in telecommunications is almost explicitly driven by essential patents and their relevant innovative technologies.[3]

All these reasons make the telecommunications sector ideal for the examination of the relationship between essential patents and standards. Examining the telecommunications sector gives us a complete insight regarding the complicated relationship between patents and standards.

As already stated, sometimes, standards containing essential patents may be developed smoothly, as in the cases of the IEEE 802.15.4 standard for a low data rate wireless personal area network and the IEEE 802.15.1 standard, which were based on already existing specifications. Sometimes, however, the standardisation process may be long and difficult as in the case of the ANSI E1-T1 standard for a very high speed digital subscriber line (VDSL), where the multi-carrier discrete multi-tone (DMT) and the single carrier technologies had a stand-off, which was a repeat of an earlier battle which had taken place during the asymmetric digital subscriber line (ADSL) negotiations. Finally, in the most severe, and rare, cases, the standardisation process may be interrupted due to lack of agreement and eventually be completely abandoned by the relevant parties, as it happened in the case of the IEEE standard 802.15.3a for a high data rate wireless personal network using ultra wideband.[11]

In contrast to what happened during the standardisation processes for ADSL

and VDSL, the ITU standard G993.2-2005 for VDSL2 was agreed upon relatively quickly,[11] which may be an indication that after a certain technology has been popularised by a prevailing standard containing essential patents regarding this technology, subsequent developments may be more easily adapted through standardisation, especially on its top level. A similar example somehow supporting this hypothesis would be Ethernet which was first patented in 1978, but only became an official IEEE 802.3 standard after it was already widely adopted as a de facto industry standard.[11]

In a similar case to the standardisation process of VDSL, different channel access methods, such as the Frequency Division Multiple Access (FDMA), the Time Division Multiple Access (TDMA) and the Code Division Multiple Access (CDMA) methods, were included in different competing second generation cellular network standards. Although these standards were initially regional, they soon started to compete against each other, with the European GSM (Global System for Mobile communications) standard, which was based on TDMA, being gradually envisioned as a worldwide one.[3][4] Therefore, the American D-AMPS (Digital Advanced Mobile Phone System) standard, which was also based on TDMA and was initially the most popular second generation cellular network standard in North America, soon after its adoption and until it was finally phased out in favour of the GSM technology, had to compete for adoption by the network carriers against both GSM and other regional American standards based on CDMA.[31]

Yet another related and interesting case is the case of the TETRA (TErrestrial Trunked RAdio) standard which is a standard for private mobile radio networks (PMR), public access mobile radio (PAMR) as well as public safety networks with application for police or fire rescue.[16] TETRA, like GSM, was adopted within ETSI, the European Telecommunications Standards Institute, and is also based on TDMA. TETRA is competitive to relevant American standards for public safety networks and neither it nor those American standards have yet completely prevailed. The standardisation process for TETRA was lengthy and included a split between the parties of the standardisation process over the channel access method selected for inclusion in the standard. Several parties left the TETRA standardisation process and formed a consortium that produced a competing standard which is called Tetrapol and is based on FDMA.[16]

An interesting detail is that Motorola, an American company, took part in the standardisation processes that resulted in both TETRA and GSM. In both cases, it was not very supportive of the idea that these standards may exceed their regional status and turn into global ones, as this could lead into Motorola losing control of the licensing policies in its home market, the United States, and also losing a significant share of the regional North American market.[3][4][16]

The case of the GSM (which initially stood for Groupe Spécial Mobile) standard is quite interesting as it was mostly formed in a newly established standardisation organisation, the European Telecommunications Standards Institute (ETSI), which did not have yet adopted comprehensive rules and guidelines concerning the inclusion of patents in standards. The different business strategies employed in the case of GSM made the need for the establishment of such rules and guidelines very evident, and it can be argued that this case made it necessary for ETSI to establish the "F/RAND obligation" for owners of patents essential for its standards, and to also adopt guidelines concerning the prior disclosure of the existence and ownership of relevant patents, and the optional ex ante disclosure of licensing terms and fees.[3][16]

In the case of the GSM standard, Motorola refused to make general declarations concerning the licensing of their relevant patents. In addition to this, while most parties involved in the standardisation process believed there was a gentleman's agreement concerning the patents of the participants which were relevant to the standard and their licensing, and did not protect their innovations and their contributions to the standard, Motorola heavily patented relevant technology while the standard was being developed. The subsequent refusal of Motorola to provide licences for its standard-essential patents unless that was done in the form of cross-licensing created barriers for all parties to enter the market, apart from those who were able to enter into a cross-licence agreement with Motorola.[3][4][16]

Only four companies managed to enter into such an agreement: Siemens, Alcatel, Nokia and Ericsson.[3][4][16] Other companies tried to secure licences from Motorola at an acceptable fee, but were unsuccessful.[3][4] In this way, Motorola tried to both secure its position in the regional European market, but also prevent the adoption of the GSM standard outside of Europe, which would have threatened its positions in other regional markets outside Europe, mainly in America.

In complete contrast to Motorola, Philips made licences for its most valuable standard-essential patents available at no cost, as its management suddenly decided to withdraw from this market.[3][4]

Motorola had also been involved in the standardisation process of the competing American standard, D-AMPS, and thus it did not agree with the gradually rising worldwide adoption of GSM, which had initiated as a European standard, not as a world-wide one. Therefore, Motorola continued its support for GSM for Europe, but simultaneously advocated other standards for use in other regions of the world.[4]

However, the end consumers' demand for a single standard in mobile telecommunications slowly drove GSM to prevail globally. While some products existed which were simultaneously compatible with both the D-AMPS and the GSM standards, most products were not. As the products that were compatible with only one of the two competing standards were already quite expensive, with royalty fees paid for the licensing of essential patents being significantly important for their prices,[8] there was an inherent demand for products compatible with a single standard and which could work and function the same both in Europe and North America. This need for compatibility led into GSM being slowly adopted worldwide.

Another incident concerning the GSM standard was the claim of a third party, the InterDigital Technology Corporation (IDC), which did not participate in the standardisation process of GSM, that its patents were infringed by GSM, and that it thus held essential patents for GSM. This claim, which was rejected in the USA, but upheld in Germany, is an example of the "non-disclosure strategy".[4]

The successor of the GSM standard, UMTS (Universal Mobile Telecommunications System), a third generation cellular network standard was adopted by an alliance, the 3rd Generation Partnership Project (3GGP), formed to ensure that all participating standardisation bodies would adopt the same standard. However a competing standard exists to UMTS, the cdma2000 standard which is based on the second generation cdmaOne standard and is mostly implemented in South Korea and the United States. The Time Division Synchronous Code Division Multiple Access (TD-SCDMA) standard, which was afterwards developed in China and implemented there was subsequently adopted by the 3GPP alliance. It has also been noted that royalty fees for standard-essential patents are still important for the price of a UMTS-compatible handset.[8]

All major regional or global standardisation organisations related to the standardisation of the telecommunications sector seem to now have adopted some form of rules concerning the inclusion of patents in standards, usually adopting "F/RAND" or similar obligations for patent owners that take part in standardisation through them,[8][18][20][32] with some of them also approving and supporting the optional ex ante disclosure of relevant patents and licensing terms and fees.[33] However, issues concerning the application of the "F/RAND obligation", its effectiveness and proposed revisions of it, concerning standardisation in general as well as the standardisation of the telecommunications sector in particular, remain open, with the International Telecommunications Union (ITU) organising a patent roundtable as late as October 2012.[34]

Furthermore, patents included in standards are very often involved in litigation between patent holders, obliging the relevant regulatory bodies to consider and review the rules and obligations imposed by standardisation organisations on patents deemed as essential for standards and included in them.[28] Most of these cases tend to concern patents included in standards related with telecommunications and other sectors with high rates of innovation.[28]

The continuous market growth of the telecommunications sector along with a constant effort to gain a bigger share of the rising total profit made in this sector have led most patent owners being involved in it and therefore also in its standardisation, to engage into simultaneous and wide-ranging court litigation and counter-litigation in multiple different regions and jurisdictions against each other over their patents and their related licensing terms and fees.[28][35][36] While litigation may not always concern standard-essential patents, it most usually does, with regulatory bodies also taking action in favour or against a certain party.[21][26][36][37]

In what has been called the "smartphone wars"[26][35][36] companies use legal action based on their patents, essential or not, to try to block each other from segments of the market, notwithstanding whether these concern geographical regions, new technological sections or various families of products. Standards and their essential patents have been employed as tools to gain a more favourable position in the market, to acquire access in technology owned by competition and, also, to prevent new competitors or others from entering the market.[3][4][29]

An odd side-effect of the current situation of patent wars is that certain companies have applied for patents on patent trolling, the practice of one engaging in constant patent litigation over one's acquired patents while having no interest in conducting research or promoting innovation.[38] This action, which could result into patent trolls having to acquire a licence to keep performing their actions, can be viewed as an effort to regulate, to control or to, even, try to dominate the field of patent litigation through indirect means, or as merely an attempt to gain another tool for defending against patent trolls.

Another interesting observation is that most cases of litigation over patents do not involve a major vendor or manufacturer against another one, but rather patent trolls, universities and individual inventors against a major vendor or manufacturer, especially concerning essential patents.[7] These cases, especially cases involving essential patents, most commonly lead to some sort of settlement or, less commonly, to the plaintiff losing the case and, very rarely to the plaintiff winning.[7]

As it has already been stated, the ownership of patents considered essential for a standard is of great value for the patent holder. This can be noted by the importance placed by Ericsson on its being the largest holder of standard-essential patents for mobile communication, while also acknowledging the significance of standardisation, cross-licensing and "F/RAND licensing".[39] The importance and value of standard-essential patents in the sector of telecommunications, and more specifically of the ones included in the most prevalent of the latest generation cellular network standards, the LTE (Long-Term Evolution) standard, has lately also been acknowledged by the Forbes magazine.[29]

Finally, it has to be recognised that new approaches must be adopted to avoid lengthy negotiations, disagreements and litigation so that patent holders can focus more on research, development and innovation.